\documentclass[11pt]{article}

\begin{document}


\hyphenation{Ra-cah-expr}
\hyphenation{Ra-cah-expr-}
\hyphenation{Ra-cah-expres-sion}

\title{Analytical expressions for special cases of $LS$--$jj$ transformation 
       matrices for a shell of equivalent electrons}

\author{G.\ Gaigalas${}^\dagger$, T.\ \v{Z}alandauskas, Z.\ Rudzikas \\
        \\
        State Institute of Theoretical Physics and Astronomy\\
        A.\ Go\v{s}tauto 12, Vilnius 2600, Lithuania.\\
        ${}^\dagger$ e--mail: gaigalas@itpa.lt \\
        \\
        \\
        } 
       
\maketitle

\begin{abstract}

Transformation matrices of the weights of the atomic wave functions in $jj$ coupling
to the relevant weights of $LS$ coupling are considered for a shell of equivalent electrons.
Their use allows one to preserve main part of relativistic effects but to classify 
the energy levels of an atom or an ion considered with the help of a set of quantum numbers 
of $LS$ coupling scheme. Having in mind that the numerical values of the abovementioned 
transformation matrix ($LS$--$jj$) may be generally obtained in a very time consuming 
recurrent way, the analytical expressions of these matrices for a number of special
cases of electronic configurations are presented.

\end{abstract}

\bigskip
\bigskip

\newpage

\section{Introduction}

It is impossible to predict, the spectroscopic data on which chemical element (neutral atom 
or its arbitrary ionization degree) will become one day urgently need. Therefore, the 
theoretical methods of the studies of the structure and properties of many-electron atoms
and ions must be fairly accurate and universal.

When identifying and classifying the energy levels of atom or ion, one has to find and use
the optimal coupling scheme of angular momenta in each open electronic shell of equivalent
electrons as well as between them. The special difficulties arise in the relativistic 
treatment of electronic configurations when the non-relativistic shell of equivalent
electrons splits into  a number of subshells. 

Offen it is important to preserve a main part of relativistic effects, but to classify 
the energy levels with the help of a set of quantum numbers of $LS$ coupling. 
This may be achieved making use of the transfomation matrices of the weights of the atomic 
functions in $jj$ coupling, obtained in relativistic approximation, to the relevant
weights of $LS$ coupling. The efficient method, implementing such a procedure, is described 
in \cite{NR,Rudzikas:97}. A number of practical applications is discussed in \cite{RC}.

Having in mind that numerical values of the transformation matrices from $jj$ to $LS$ coupling
needed may be obtained generally in very time consuming recurrent way, the possibilities 
to find the analytical expressions of such matrices for a number of special cases 
of electronic configurations are discussed in this paper.


\section{General consideration of the $LS$--$jj$ transformation matrices for a shell of 
equivalent electrons $l^{N}$}

The relativistic analogue of the non-relativistic wave function of a shell of $l^{N}$ 
equivalent electrons ($LS$ coupling)  

\begin{equation}
\label{LSstate}
   |l^N \alpha LS),
\end{equation}
where $\alpha$ denotes the additional quantum numbers necessary for the one-to-one classification  
of the energy levels of that shell, in the $jj$ coupling splits into so-called subshells. 
It will look as follows:

\begin{equation}
\label{jjSatte}
   |lj_{1}^{N_{1}}j_{2}^{N_{2}}\nu_{1}J_{1}\nu_{2}J_{2}J),
\end{equation}
where $j_{1}=l-\frac{1}{2}$, $j_{2}=l+\frac{1}{2}$ and  $N_{1}+N_{2}=N$. For subshells with 
angular momenta $j_{i}=\frac{1}{2},$ $\frac{3}{2},$ $\frac{5}{2}$ and $\frac{7}{2}$ (this 
corresponds to $s^{N}$, $p^{N}$, $d^{N}$ and $f^{N}$ shells as well as  $j = l-\frac{1}{2}$ 
of $g^{N}$ shell) two quantum numbers $\nu$ and $J$ are sufficient to unabiguously classify 
the relevant states.

\medskip

The following relationship between the weights $c_{ik}$ and $a_{jr}$ of the wave 
functions of two pure coupling schemes is known \cite{NR,Rudzikas:97}: 
\begin{equation}
\label{Matrix}
   c_{jk} = \sum_{r}a_{jr}(\phi_{k}|\psi_{r}).
\end{equation}

\medskip
     
It may be used to transform the weights of the relativistic wave function in $jj$ coupling to
the relevant weights of $LS$ coupling. However, in such a case this equality 
becomes not completely accurate bacause of the changes of the character of the wave functions and the
Hamiltonian (non-relativistic and relativistic). Actually a part of the relativistic
effects is lost after this transformation. Further on we shall discuss the transformation 
matrix $(\phi_{k}|\psi_{r})$ in Eq (\ref{Matrix}).
  
The $LS$--$jj$ transformation matrix 
$(l^{N}\alpha LSJ|lj_{1}^{N_{1}}j_{2}^{N_{2}}\nu_{1}J_{1}\nu_{2}J_{2}J)$ 
connecting two wave functions of a shell of equivalent electrons, namely
$|l^{N}\alpha LSJ)$ in $LS$ coupling and 
$|lj_{1}^{N_{1}}j_{2}^{N_{2}}\nu_{1}J_{1}\nu_{2}J_{2}J)$
in $jj$ coupling is defined as follows:

\begin{eqnarray}
\label{MatrixLS-jjDef}
   |lj_{1}^{N_{1}}j_{2}^{N_{2}}\nu_{1}J_{1}\nu_{2}J_{2}J) = 
\sum_{\alpha LS}{|l^{N}\alpha LSJ) 
(l^{N}\alpha LSJ|lj_{1}^{N_{1}}j_{2}^{N_{2}}\nu_{1}J_{1}\nu_{2}J_{2}J)}
\end{eqnarray}
and 
\begin{eqnarray}
\label{Matrixjj-LSDef}
   |l^{N}\alpha LSJ) = \sum_{\nu_{1}J_{1}\nu_{2}J_{2}N_{1}}
{|lj_{1}^{N_{1}}j_{2}^{N_{2}}\nu_{1}J_{1}\nu_{2}J_{2}J)}
(lj_{1}^{N_{1}}j_{2}^{N_{2}}\nu_{1}J_{1}\nu_{2}J_{2}J|l^{N}\alpha LSJ ).
\end{eqnarray}

\medskip

The phase system of the wave functions is usually chosen in such a way that the 
coefficients of 
fractional parentage (CFP) are real numbers. Then the $LS$--$jj$ transformation matrices 
will be real too. In the case of orthonormal wave functions the transformation matrices 
will also obey the following orthonormality relations:

\begin{eqnarray}
\label{Ortogonalityjj}
   \sum_{\alpha LS}
(lj_{1}^{N_{1}^{\prime}}j_{2}^{N_{2}^{\prime}}\nu_{1}^{\prime}J_{1}^{\prime}\nu_{2}^{\prime}J_{2}
^{\prime}J | l^{N}\alpha LSJ )
(l^{N}\alpha LSJ|lj_{1}^{N_{1}}j_{2}^{N_{2}}\nu_{1}J_{1}\nu_{2}J_{2}J)  
\nonumber \\[0.2cm]
= \delta
(N_{1}^{\prime}N_{2}^{\prime}\nu_{1}^{\prime}J_{1}^{\prime}\nu_{2}^{\prime}J_{2}^{\prime},
N_{1}N_{2}\nu_{1}J_{1}\nu_{2}J_{2})
\end{eqnarray}
   
and

\begin{eqnarray}
\label{OrtogonalityLS}
   \sum_{\nu_{1}J_{1}\nu_{2}J_{2}N_{1}}
(l^{N}\alpha^{\prime} L^{\prime}S^{\prime}J|lj_{1}^{N_{1}}j_{2}^{N_{2}}\nu_{1}J_{1}\nu_{2}J_{2}J)
(lj_{1}^{N_{1}}j_{2}^{N_{2}}\nu_{1}J_{1}\nu_{2}J_{2}J|l^{N}\alpha LSJ ) = 
\delta (\alpha^{\prime} L^{\prime}S^{\prime},\alpha LS).
\end{eqnarray}


Transformation matrices considered may be calculated using the following 
reccurent relation \cite{Rudzikas:97}:

\begin{eqnarray}
\label{Main}
(l^{N}\alpha LSJ|lj_{1}^{N_{1}}j_{2}^{N_{2}}\nu_{1}J_{1}\nu_{2}J_{2}J) =
\sqrt{\left[L,S\right]/N} \sum_{\alpha^{\prime} L^{\prime}S^{\prime}} 
(l^{N}\alpha LS||l^{N-1}(\alpha^{\prime} L^{\prime}S^{\prime})l)
\sum_{J^{\prime}}\left[J^{\prime}\right]
\nonumber \\[0.2cm]
\times~
\left[
\sqrt{N_{1}\left[j_{1},J_{1}\right]}
\left\{
\begin{array}{ccc}
   L^{\prime} & l & L \\
   S^{\prime} & s & S \\
   J^{\prime} & j_{1} & J
\end{array}
\right\}
\sum_{\nu_{1}^{\prime}J_{1}^{\prime}}(-1)^{j_{1}+J_{1}-J_{2}+J^{\prime}}
\left\{
\begin{array}{ccc}
   J_{2} & J_{1}^{\prime} & J^{\prime} \\
   j_{1} & J & J_{1}
\end{array}
\right\}
\right. 
\nonumber \\[0.2cm]
\left.
\times~
(j_{1}^{N_{1}-1}(\nu_{1}^{\prime}J_{1}^{\prime})j_{1}||j_{1}^{N_{1}}\nu_{1}J_{1})
(l^{N-1}\alpha^{\prime}L^{\prime}S^{\prime}J^{\prime} |
lj_{1}^{N_{1}-1}j_{2}^{N_{2}}\nu_{1}^{\prime}J_{1}^{\prime}\nu_{2}J_{2}J^{\prime})
\right. 
\nonumber \\[0.2cm]
\left.
+~
\sqrt{N_{2}\left[j_{2},J_{2}\right]}
\left\{
\begin{array}{ccc}
   L^{\prime} & l & L \\
   S^{\prime} & s & S \\
   J^{\prime} & j_{2} & J
\end{array}
\right\}
\sum_{\nu_{2}^{\prime}J_{2}^{\prime}}(-1)^{j_{2}+J_{1}+J_{2}^{\prime}+J}
\left\{
\begin{array}{ccc}
   J_{2}^{\prime} & J_{1} & J^{\prime} \\
   J & j_{2} & J_{2}
\end{array}
\right\}
\right. 
\nonumber \\[0.2cm]
\left.
\times~
(j_{2}^{N_{2}-1}(\nu_{2}^{\prime}J_{2}^{\prime})j_{2}||j_{2}^{N_{2}}\nu_{2}J_{2})
(l^{N-1}\alpha^{\prime}L^{\prime}S^{\prime}J^{\prime} |
lj_{1}^{N_{1}}j_{2}^{N_{2}-1}\nu_{1}J_{1}\nu_{2}^{\prime}J_{2}^{\prime}J^{\prime})
\right]
\end{eqnarray}

starting with

\begin{eqnarray}
\label{KrS1}
(l^{2}LSJ|lj_{1}j_{2}J) = \frac{1}{\sqrt{2}}\left(1+(-
1)^{L+S}\right)\sqrt{\left[j_{1},j_{2},L,S\right]}
\left\{
\begin{array}{ccc}
   l & l & L \\
   s & s & S \\
   j_{1} & j_{2} & J
\end{array}
\right\},
\end{eqnarray}

and 

\begin{eqnarray}
\label{KrS2}
(l^{2}LSJ|lj^{2}J) = \frac{1}{4}\left(1+(-1)^{L+S}\right)\left(1+(-
1)^{J}\right)\left[j\right]\sqrt{\left[L,S\right]}
\left\{
\begin{array}{ccc}
   l & l & L \\
   s & s & S \\
   j & j & J
\end{array}
\right\}.
\end{eqnarray}

\medskip

Here $\left[a,b\right]=(2a+1)(2b+1)$.\\
The equation (\ref{Main}) requires the CFP in $LS$ and $jj$ couplings, namely  
$(l^{N}\alpha LS||l^{N-1}(\alpha^{\prime} L^{\prime}S^{\prime})l)$ and
$(j^{N-1}(\nu^{\prime}J^{\prime})j||j^{N}\nu J)$. It is efficient to use the CFP obtained from reduced 
coefficients of fractional parentage (RCFP) \cite{Rudzikas:97}. The complete tables of their 
numerical values (the $f^{N}$ shells included) may be found in \cite{RCFP_LS,RCFP_jj}. 
This gives us natural relation between the 
CFP's of partially and almost filled shells:   
\begin{eqnarray}
\label{RelCFPLS}
(l^{4l+1-N}(\alpha^{\prime} \nu^{\prime}L^{\prime}S^{\prime})l||l^{4l+2-N}\alpha \nu LS) = 
(-1)^{S+S^{\prime}+L+L^{\prime}-l-\frac{1}{2}+\frac{1}{2}(\nu + \nu^{\prime}-1)}
\nonumber \\[0.2cm]
\times 
\left(
\frac{(N+1)(2L^{\prime}+1)(2S^{\prime}+1)}{(4l+2-N)(2L+1)(2S+1)}
\right)^{\frac{1}{2}}
(l^{N}(\alpha^{\prime} \nu^{\prime}L^{\prime}S^{\prime})l||l^{N+1}\alpha \nu LS),
\end{eqnarray}

\begin{eqnarray}
\label{RelCFPjj}
(j^{2j-N}(\nu^{\prime}J^{\prime})j||j^{2j+1-N}\nu J) 
\nonumber \\[0.2cm]
=(-1)^{J+J^{\prime}-j+\frac{1}{2}(\nu+\nu^{\prime}-1)}
\left(
\frac{(N+1)(2J^{\prime}+1)}{(2j+1-N)(2J+1)}
\right)^{\frac{1}{2}}
(j^{N}(\nu^{\prime}J^{\prime})j||j^{N+1}\nu J).
\end{eqnarray}

\medskip

As noticed in \cite{Grant} the $LS$--$jj$ transformation matrices calculated using CFP's which 
satisfy the conditions (\ref{RelCFPLS}) and (\ref{RelCFPjj}) are related by the following simple 
symmetry property:
\begin{eqnarray}
\label{RelLSjj}
(l^{N}\alpha\nu LSJ|lj_{1}^{N_{1}}j_{2}^{N_{2}}\nu_{1}J_{1}\nu_{2}J_{2}J) 
\nonumber \\[0.2cm] 
=(-1)^{(\nu-\nu_{1}-\nu_{2})/2}
(l^{4l+2-N}\alpha\nu LSJ|
lj_{1}^{2j_{1}+1-N_{1}}j_{2}^{2j_{2}+1-N_{2}}\nu_{1}J_{1}\nu_{2}J_{2}J).
\end{eqnarray}
Therefore it is usually enough to calculate using Eq. (\ref{Main})
only the matrix elements of partially filled $l^{N}$ shells.  

\medskip


\section{Special cases of transformation matrices}

The explicit formulas for such matrices in the case of configurations $l^{N}$ and 
$l_{1}^{N_{1}}l_{2}^{N_{2}}$ (including the special case of configurations 
$p^{N}l_{2}^{N_{2}}$ and $l_{1}^{N_{1}}p^{N}$) are considered in \cite{KCSSR}. Below 
let us consider the possibilities to obtain analytical expressions for such transformation
matrices.

In \cite{Rudzikas:97} there are presented algebraic expressions for a fairly large 
number of particular cases of CFP. Making use of them we could find similar algebraic 
formulas for the transformation matrices under consideration.
Let us discuss the case of $l^{3}$ as the example.
Further on we will investigate the cases when in one of $jj$ subshells there is 1 electron 
and in other 2 electrons (we will denote this case as 
$N_{1} = 1$, $N_{2} = 2$ or $N_{1} = 2$, $N_{2} = 1$) and 
the case when all three electrons are in one subshell ($N_{1} = 3$ or $N_{2} = 3$).
   
\medskip


{\bf a)} $N_{1} = 1$, $N_{2} = 2$ or $N_{1} = 2$, $N_{2} = 1$

Let us assume that $j_{1}= l - \frac{1}{2}$ and $j_{2}= l + \frac{1}{2}$.
Inserting the expressions for transformation matrix of a shell having two 
electrons, namely (\ref{KrS1}),(\ref{KrS2}) and known expressions for CFP 
(for example, (9.14) from \cite{Rudzikas:97}) into (\ref{Main}) we get: 

\begin{eqnarray}
\label{l3_12_big}
(l^{3}\alpha LSJ|lj_{1}^{1}j_{2}^{2}\nu_{2}J_{2}\nu J) = 
\frac{1}{2\sqrt{3}}(1+(-1)^{J_{2}})\left[j_{2}\right]
\sqrt{\left[L,S,j_{1},J_{2}\right]}
\times
\nonumber \\[0.2cm]
\times
\sum_{\alpha^{\prime}L^{\prime}S^{\prime}}
(l^{3} \alpha L S || l^{2}(\alpha^{\prime}L^{\prime}S^{\prime})l)
\sqrt{\left[L^{\prime},S^{\prime}\right]}(1+(-1)^{L^{\prime}+S^{\prime}})
\times
\nonumber \\[0.2cm]
\times
\left[
\frac{1}{2}
\left\{
\begin{array}{ccc}
   l & l & L^{\prime} \\
   s & s & S^{\prime} \\
   j_{2} & j_{2} & J_{2}
\end{array}
\right\}
\left\{
\begin{array}{ccc}
   l & L & L^{\prime} \\
   s & S & S^{\prime} \\
   j_{1} & J & J_{2}
\end{array}
\right\}
+
\sum_{J^{\prime}} \left[J^{\prime}\right]
\left\{
\begin{array}{ccc}
   j_{2} & j_{1} & J^{\prime} \\
   J & j_{2} & J_{2}
\end{array}
\right\}
\left\{
\begin{array}{ccc}
   l & l & L^{\prime} \\
   s & s & S^{\prime} \\
   j_{1} & j_{2} & J^{\prime}
\end{array}
\right\}
\left\{
\begin{array}{ccc}
   l & L & L^{\prime} \\
   s & S & S^{\prime} \\
   j_{2} & J & J^{\prime}
\end{array}
\right\}
\right]
\end{eqnarray}

\medskip

Further simplifications are not obvious. Nevertheless we found that 

\begin{eqnarray}
\label{l3_12_Prop}
%
\sum_{\alpha^{\prime}L^{\prime}S^{\prime}}
(l^{3} \alpha L S || l^{2}(\alpha^{\prime}L^{\prime}S^{\prime})l)
\sqrt{\left[L^{\prime},S^{\prime}\right]}(1+(-1)^{L^{\prime}+S^{\prime}})
\left\{
\begin{array}{ccc}
   l & l & L^{\prime} \\
   s & s & S^{\prime} \\
   j_{2} & j_{2} & J_{2}
\end{array}
\right\}
\left\{
\begin{array}{ccc}
   l & L & L^{\prime} \\
   s & S & S^{\prime} \\
   j_{1} & J & J_{2}
\end{array}
\right\}
\nonumber \\[0.2cm]
=
%
\sum_{\alpha^{\prime}L^{\prime}S^{\prime}}
(l^{3} \alpha L S || l^{2}(\alpha^{\prime}L^{\prime}S^{\prime})l)
\sqrt{\left[L^{\prime},S^{\prime}\right]}(1+(-1)^{L^{\prime}+S^{\prime}})
\nonumber \\[0.2cm]
\times
\sum_{J^{\prime}} \left[J^{\prime}\right]
\left\{
\begin{array}{ccc}
   j_{2} & j_{1} & J^{\prime} \\
   J & j_{2} & J_{2}
\end{array}
\right\}
\left\{
\begin{array}{ccc}
   l & l & L^{\prime} \\
   s & s & S^{\prime} \\
   j_{1} & j_{2} & J^{\prime}
\end{array}
\right\}
\left\{
\begin{array}{ccc}
   l & L & L^{\prime} \\
   s & S & S^{\prime} \\
   j_{2} & J & J^{\prime}
\end{array}
\right\}.
\end{eqnarray}

So, using (\ref{l3_12_Prop}) we get the following simplified expression of matrix elements 
for the case $N_{1} = 1$, $N_{2} = 2$: 

\begin{eqnarray}
\label{l3_12}
(l^{3}\alpha LSJ|lj_{1}^{1}j_{2}^{2}\nu_{2}J_{2} J) = 
\frac{\sqrt{3}}{4}(1+(-1)^{J_{2}})\left[j_{2}\right]
\sqrt{\left[L,S,j_{1},J_{2}\right]}
\nonumber \\[0.2cm]
\times
\sum_{\alpha^{\prime}L^{\prime}S^{\prime}}
(l^{3} \alpha L S || l^{2}(\alpha^{\prime}L^{\prime}S^{\prime})l)
\sqrt{\left[L^{\prime},S^{\prime}\right]}(1+(-1)^{L^{\prime}+S^{\prime}})
\left\{
\begin{array}{ccc}
   l & l & L^{\prime} \\
   s & s & S^{\prime} \\
   j_{2} & j_{2} & J_{2}
\end{array}
\right\}
\left\{
\begin{array}{ccc}
   l & L & L^{\prime} \\
   s & S & S^{\prime} \\
   j_{1} & J & J_{2}
\end{array}
\right\}.
\end{eqnarray}

One can obtain the relevant expression of transformation matrix elements for the case 
$N_{1} = 2$, $N_{2} = 1$ from (\ref{l3_12}) transposing $j_{1}\longleftrightarrow j_{2}$ 
and $J_{2} \rightarrow J_{1}$. 
The simplified expression (\ref{l3_12}) was used to check numerically the general
formulae (\ref{Main}) for all cases of $l^{3}$, $l$=1,2 and 3.

\medskip


{\bf b)} $N_{1} = 3$, $N_{2} = 0$ or $N_{1} = 0$, $N_{2} = 3$ 

Let us now consider the case when all three electrons are in one $jj$ subshell, namely $j^{3}$. 
Inserting the expressions of the transfomation matrices for a shell of two 
equivalent electrons into (\ref{Main}) we obtain: 

\begin{eqnarray}
\label{l3_30_big}
(l^{3}\alpha LSJ|lj^{3}\nu J) = 
\frac{1}{4}\sqrt{\left[L,S\right]}\left[j\right]^{\frac{3}{2}} 
\sum_{\alpha^{\prime}L^{\prime}S^{\prime}}
(l^{3} \alpha L S || l^{2}(\alpha^{\prime}L^{\prime}S^{\prime})l)
\sqrt{\left[L^{\prime},S^{\prime}\right]}
(1+(-1)^{L^{\prime}+S^{\prime}})
\times
\nonumber \\[0.2cm]
\times 
\sum_{J^{\prime}} \sqrt{\left[J^{\prime}\right]}
(1+(-1)^{J^{\prime}})
\left\{
\begin{array}{ccc}
   l & l & L^{\prime} \\
   s & s & S^{\prime} \\
   j & j & J^{\prime}
\end{array}
\right\}
\left\{
\begin{array}{ccc}
   l & L & L^{\prime} \\
   s & S & S^{\prime} \\
   j & J & J^{\prime}
\end{array}
\right\}
(j^{2}(J^{\prime})j||j^{3}\nu J).
\end{eqnarray}


For $(j^{2}(J^{\prime})j||j^{3} J)$ we use analytical expression 


\begin{eqnarray}
\label{CFP_jj}
(j^{2}(J^{\prime}),j||j^{3}J) = 
\frac{(-1)^{j-J}}{\sqrt{1+2\left[J_{0}\right]
\left\{
\begin{array}{ccc}
   j & j & J_{0} \\
   j & J & J_{0}
\end{array}
\right\}
}}
\frac{(1+(-1)^{J_{0}})}{2\sqrt{3}}
\nonumber \\[0.2cm]
\left[
\delta(J^{\prime},J_{0})\delta(jjJ_{0})\delta(jJ_{0}J) + 
(1+(-1)^{J^{\prime}})\sqrt{\left[J_{0},J^{\prime}\right]}
\left\{
\begin{array}{ccc}
   j & j & J^{\prime} \\
   j & J & J_{0}
\end{array}
\right\}
\right],
\end{eqnarray}

where parameter $J_{0}$ should be chosen to satisfy two conditions: triangular condition
$\delta(jJ_{0}J)$ should be satisfied and $J_{0}$ should be even. 
The expression (\ref{CFP_jj}) is valid when the quantum number $J$ is sufficient to classify 
the state in $jj$ coupling.

\medskip

Inserting (\ref{CFP_jj}) into (\ref{l3_30_big}) we arrive at the expressions similar to 
(\ref{l3_12_big}). 
Then analogically to (\ref{l3_12_Prop}) we found the property 

\begin{eqnarray}
\label{l3_30_Prop}
\sum_{\alpha^{\prime}L^{\prime}S^{\prime}}
(l^{3} \alpha L S || l^{2}(\alpha^{\prime}L^{\prime}S^{\prime})l)
\sqrt{\left[L^{\prime},S^{\prime}\right]}(1+(-1)^{L^{\prime}+S^{\prime}})
\left\{
\begin{array}{ccc}
   l & l & L^{\prime} \\
   s & s & S^{\prime} \\
   j & j & J_{0}
\end{array}
\right\}
\left\{
\begin{array}{ccc}
   l & L & L^{\prime} \\
   s & S & S^{\prime} \\
   j & J & J_{0}
\end{array}
\right\}
\nonumber \\[0.2cm]
=
\sum_{\alpha^{\prime}L^{\prime}S^{\prime}}
(l^{3} \alpha L S || l^{2}(\alpha^{\prime}L^{\prime}S^{\prime})l)
\sqrt{\left[L^{\prime},S^{\prime}\right]}(1+(-1)^{L^{\prime}+S^{\prime}})
\nonumber \\[0.2cm]
\times
\sum_{J^{\prime}} \left[J^{\prime}\right]
\left\{
\begin{array}{ccc}
   j & j & J^{\prime} \\
   J & j & J_{0}
\end{array}
\right\}
\left\{
\begin{array}{ccc}
   l & l & L^{\prime} \\
   s & s & S^{\prime} \\
   j & j & J^{\prime}
\end{array}
\right\}
\left\{
\begin{array}{ccc}
   l & L & L^{\prime} \\
   s & S & S^{\prime} \\
   j & J & J^{\prime}
\end{array}
\right\}.
\end{eqnarray}

Using (\ref{l3_30_Prop}) we find the following simplified expressions for the case of 
$N_{i} = 3$:

\begin{eqnarray}
\label{l3_30}
(l^{3}\alpha LSJ|lj^{3}\nu J) = 
\frac{\sqrt{3}}{2}\sqrt{\left[L,S,J_{0}\right]}\left[j\right]^{\frac{3}{2}}
\frac{(-1)^{j-J}}{\sqrt{1+2\left[J_{0}\right]
\left\{
\begin{array}{ccc}
   j & j & J_{0} \\
   j & J & J_{0}
\end{array}
\right\}
}}
\nonumber \\[0.2cm]
\times
\sum_{\alpha^{\prime}L^{\prime}S^{\prime}}
(l^{3} \alpha L S || l^{2}(\alpha^{\prime}L^{\prime}S^{\prime})l)
\sqrt{\left[L^{\prime},S^{\prime}\right]}(1+(-1)^{L^{\prime}+S^{\prime}})
\left\{
\begin{array}{ccc}
   l & l & L^{\prime} \\
   s & s & S^{\prime} \\
   j & j & J_{0}
\end{array}
\right\}
\left\{
\begin{array}{ccc}
   l & L & L^{\prime} \\
   s & S & S^{\prime} \\
   j & J & J_{0}
\end{array}
\right\}.
\end{eqnarray}

\medskip

Formulas like Eq. (\ref{l3_12}) and (\ref{l3_30}) are simplier for calculations compared to 
general recurrent relation (\ref{Main}). 
We can even more simplify such formulas with the use of known algebraic expressions for 
CFP's in $LS$ and $jj$ couplings.
Such expressions and the relevant computer code may be useful as an independent 
check of more general programs for $LS$--$jj$ matrix elements calculation.
Similar expressions may be found for a number of other special cases of electronic 
configurations, but they are rather complicated and, therefore, of a little use.

\section{Conclusion}

Numerical values of the transformation matrices from $jj$ to $LS$ coupling scheme in 
general may be found making use of the recurrent formulae of the kind (\ref{Main}).
The relevant calculation procedure is very time consuming, therefore the alternative 
ways, even if they are suitable only for some particular cases, are of interest. 

The use of analytical expressions for the coefficients of fractional parentage,
both in $LS$ and $jj$ couplings, allowed us to obtain the comparatively simple algebraic
formulas for the abovementioned transformation matrices in the case of particular electronic
configurations.


\end{document}